\documentclass[a4paper,11pt]{article}
\usepackage{pos}
\usepackage{subcaption}
\usepackage{braket}
\usepackage[font=small, labelfont=bf, textfont={small,it}]{caption}
\graphicspath{{./}}

\newcommand{\Tr}{\mathrm{Tr}}
\newcommand{\Var}{\mathrm{Var}}
\newcommand{\CP}{\mathrm{CP}}
\newcommand{\SU}{\mathrm{SU}}
\newcommand{\Sp}{\mathrm{Sp}}
\newcommand{\dd}{\mathrm{d}}
\newcommand{\nstep}{n_{\scriptscriptstyle{\rm step}}}
\newcommand{\nbet}{n_{\scriptscriptstyle{\rm between}}}
\newcommand{\tauint}{\tau_{\scriptscriptstyle{\rm int}}}
\newcommand{\DKL}{\widetilde{D}_{\scriptscriptstyle{\rm KL}}}
\newcommand{\ESS}{\mathrm{ESS}}
\newcommand{\NE}{{\scriptscriptstyle{\rm NE}}}
\newcommand{\f}{{\scriptscriptstyle{\rm f}}}
\renewcommand{\r}{{\scriptscriptstyle{\rm r}}}
\newcommand{\beq}{\begin{eqnarray}}
\newcommand{\eeq}{\end{eqnarray}}

\title{Topological susceptibility of $\SU(3)$ pure-gauge theory from out-of-equilibrium simulations}
\ShortTitle{The $\SU(3)$ topological susceptibility from out-of-equilibrium simulations}

\author[a]{Claudio Bonanno}
\author[b]{Alessandro Nada}
\author*[c]{Davide Vadacchino}
\affiliation[a]{Instituto de F\'isica T\'eorica UAM-CSIC, c/ Nicol\'as Cabrera 13-15, Universidad Aut\'onoma de Madrid, Cantoblanco, E-28049 Madrid, Spain}
\affiliation[b]{Dipartimento di Fisica,  Universit\'a degli Studi di Torino and INFN, Sezione di Torino, Via Pietro Giuria 1, I-10125 Turin, Italy}
\affiliation[c]{Centre for Mathematical Sciences, University of Plymouth, Plymouth, PL4 8AA, United Kingdom}

\emailAdd{claudio.bonanno@csic.es}
\emailAdd{alessandro.nada@unito.it}
\emailAdd{davide.vadacchino@plymouth.ac.uk}

\abstract{In \textit{JHEP} \textbf{04} (2024) 126 [arXiv:2402.06561] we recently proposed an out-of-equilibrium setup to reduce the large auto-correlations of the topological charge in two-dimensional $\mathrm{CP}^{N-1}$ models. Our proposal consists of performing open-boundaries simulations at equilibrium, and gradually switching on periodic boundary conditions out-of-equilibrium. Our setup allows to exploit the reduced auto-correlations achieved with open boundaries, avoiding at the same time unphysical boundary effects thanks to a Jarzynski-inspired reweighting-like procedure. We present preliminary results obtained applying this setup to the $4d$ $\mathrm{SU}(3)$ pure-gauge theory and we outline a computational strategy to mitigate topological freezing in this theory.}

\FullConference{The 41$^{\text{st}}$ International Symposium on Lattice Field Theory (LATTICE2024)\\
 28 July -- 3 August 2024\\
Liverpool, UK\\}

\begin{document}
\maketitle

\section{Introduction}

Over the last 50 years, lattice field theory has proven to be a successful
first-principle approach to the study of the non-perturbative regime of
strongly-interacting gauge theories and in particular Quantum Chromo--Dynamics (QCD). 
Crucial to this success have been the striking
advances made in the study and design of algorithms employed to explore
their phase space.

\begin{figure}[!t]
\centering
\includegraphics[scale=0.5]{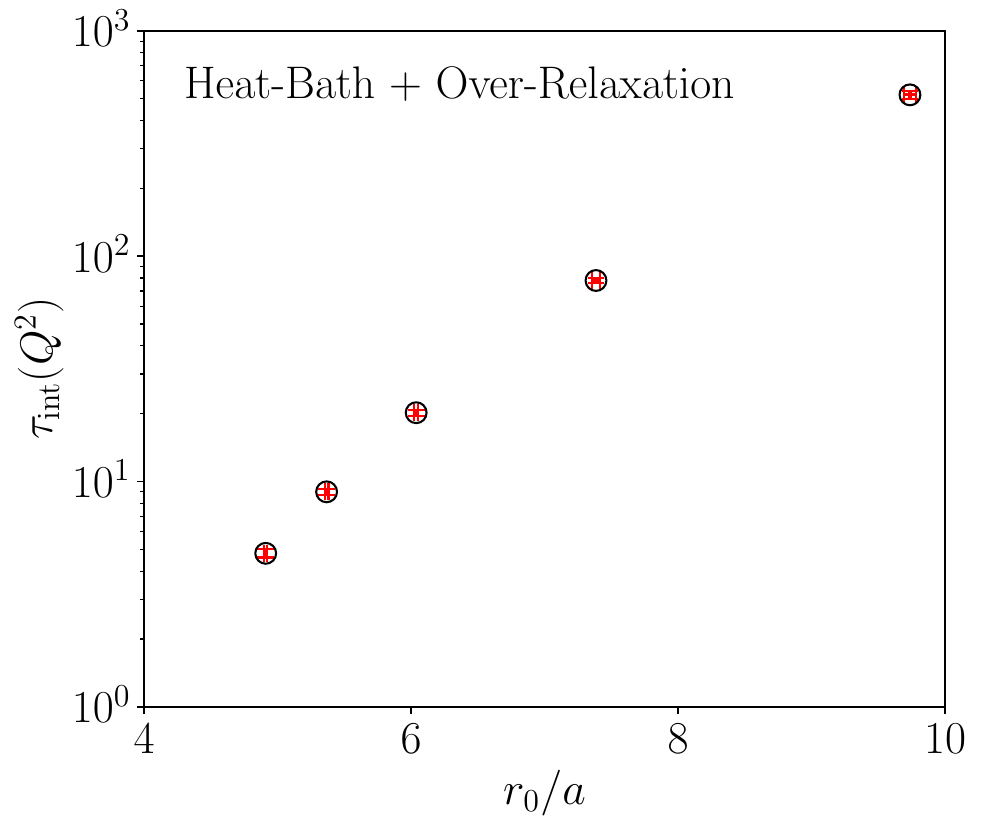}
\caption{Growth of the integrated auto-correlation time $\tauint$ of $Q^2$ as a function of the inverse lattice spacing. This quantity expresses the number of updating steps required to generate two decorrelated measures of $Q^2$. In this case $\tauint(Q^2)$ is given in units of standard updating steps, defined as 4 lattice sweeps of over-relaxation and 1 lattice sweep of heat-bath. Data for $\tauint(Q^2)$ are taken from dataset generated in Ref.~\cite{Bonanno:2023ple}, while the scale setting in units of the Sommer parameter $r_0\simeq 0.5$ fm is taken from the interpolation of the results of Ref.~\cite{Necco:2001xg}.}
\label{fig:freezing}
\end{figure}

Despite these advances, the simulation of theories characterized by
topologically non-trivial phase spaces 
close to the continuum limit remains difficult. 
This state of affairs is caused by 
\emph{topological freezing}, a severe divergence of the integrated
correlation time $\tauint$ of the the topological charge as a function of the 
correlation length in the approach towards the continuum limit of
said theories. In contrast to other cases of critical slowing down for non-topological quantities (e.g., the Wilson loop), 
where the divergence is polynomial in the lattice correlation length $\xi_{\scriptscriptstyle{\rm L}}$ with a small exponent (typically $\tauint\sim \xi_{\scriptscriptstyle{\rm L}}^z$ with $z\simeq2$ in the continuum limit $\xi_{\scriptscriptstyle{\rm L}}\to\infty$), topological freezing is characterized
by a much more dramatic growth. This has been known to affect both $\SU(N)$ and
$\Sp(2N)$ Yang--Mills gauge theories in four dimensions, as well as other lower-dimensional Quantum Field Theories whose vacuum state possess non-trivial topological features, see Refs.~\cite{Alles:1996vn,DelDebbio:2004xh,Bonati:2017woi,
Bennett:2022ftz} and Fig.~\ref{fig:freezing}.
In practice, this causes the Markov Chain Monte Carlo (MCMC) trajectory
of the system to remain trapped in a sector of the phase space with fixed topological charge, and this loss of ergodicity can potentially introduce unwanted biases in the estimation of topological and non-topological physical observables.

Given the theoretical and phenomenological relevance of observables like the topological susceptibility both for QCD hadron phenomenology~\cite{Ce:2015qha,Ce:2016awn,Bonati:2015sqt,Bonati:2016tvi,Bonanno:2020hht,Athenodorou:2020ani,Athenodorou:2021qvs,Bonanno:2023ple} and for Beyond Standard 
Model physics~\cite{Bonati:2015vqz,Petreczky:2016vrs,Borsanyi:2016ksw,Lombardo:2020bvn,Bennett:2022ftz,Athenodorou:2022aay}, several proposal have been put forward over the years in the
attempt
of addressing the issue of topological freezing, see Refs.~\cite{Bietenholz:2015rsa,Laio:2015era,Luscher:2017cjh,Hasenbusch:2017unr,Bonati:2017woi,Giusti:2018cmp,Florio:2019nte,Funcke:2019zna,Bonanno:2020hht,Kanwar:2020xzo,Nicoli:2020njz,Albandea:2021lvl,Cossu:2021bgn,Borsanyi:2021gqg,papamakarios2021,Fritzsch:2021klm,Abbott:2023thq,Eichhorn:2023uge,Howarth:2023bwk,Albandea:2024fui,Bonanno:2024zyn},
and Refs.~\cite{Finkenrath:2023sjg,Boyle:2024nlh,Finkenrath:2024ptc} for
recent reviews. 

A popular strategy in QCD simulations with dynamical fermions consists in performing simulations with Open Boundary Conditions on the temporal side of the lattice~\cite{Luscher:2011kk,Luscher:2012av}. This approach reduces the severity of topological critical slowing down, but at the cost of enhancing finite size effects, which manifest in the form of boundary effects. Another recent strategy that has been extensively applied both in pure-gauge theories~\cite{Hasenbusch:2017unr,Berni:2019bch,Bonanno:2020hht,Bonanno:2022hmz,Bonanno:2022yjr,DasilvaGolan:2023cjw,Bonanno:2023hhp,Bonanno:2024ggk,Bonanno:2024nba} and in the presence of dynamical fermions~\cite{Bonanno:2024zyn} combines Periodic and Open Boundary Conditions in a parallel tempering framework (Parallel Tempering on Boundary Conditions---PTBC) in order to avoid systematic boundary effects. 
In Ref.~\cite{Bonanno:2024udh} we have recently proposed a novel approach designed to mitigate topological freezing which shares a few similarities with PTBC. This new setup is rooted on Jarzynski's equality~\cite{Jarzynski:1996oqb} and its application to non-equilibrium
evolutions in lattice field theory (see Refs.~\cite{Caselle:2016wsw,Caselle:2018kap,Francesconi:2020fgi,Bulgarelli:2023ofi}) to combine Open and Periodic Boundary Conditions.

In this contribution, we apply our out-of-equilibrium approach to the four dimensional
$\SU(3)$ quenched lattice gauge theory. Our aim is to perform a preliminary
assessment of the viability of our strategy by measuring the topological
susceptibility $\chi$ and the integrated autocorrelation time of the topological
charge $\tauint(Q^2)$ at a moderately small value of the lattice spacing.
This enables us to perform a comparison with traditional approaches, 
that are still viable in this regime, and with other recent state-of-the-art
algorithms, like the Parallel Tempering on Boundary
Conditions. 
Moreover, this affords us a clear understanding of the computational
cost of our approach. 
This is the first step towards undertaking the same
analysis closer to the continuum limit, where traditional approaches are
doomed to fail in sampling non-trivial topology. This also provides a first
solid test-bed to implement improvements based on recent machine-learning
techniques like Stochastic Normalizing Flows, see for example Refs.~\cite{wu2020stochastic,Caselle:2022acb,Bulgarelli:2024cqc}, 
that fit into this approach quite naturally.

This proceeding is structured as follows. In Sec.~2 we briefly describe
our approach and provide details about our setup. In Sec.~3 we 
report on our numerical results and discuss them. We then
conclude and provide a roadmap for the future steps of this analysis in Sec.~4.

\section{Setup}

Our approach it based on the use of non-equilibrium evolutions and Jarzynski's
inequality to resample ensembles of configurations produced with Open
Boundary Conditions (on a defect).
It is well known that the use of Open Boundary Conditions (OBC) strongly mitigates topological freezing. The crucial difference with Periodic Boundary
Conditions (PBC) is the absence, in the former case, of those potential
barriers that separate different topological sectors, causing the freezing
of topological charge in the first place. While effective in overcoming
topological freezing, OBC come with the important drawback of introducing
non-physical effects that need to be removed by computing correlation functions sufficiently far from the boundaries. 
Our strategy is designed to exploit the advantages of OBC without suffering from its pitfalls (similarly to the PTBC algorithm).

First, the so-called defect is defined: a three-dimensional subset of the 
lattice on which the coupling can be tuned with the purpose of interpolating
between OBC, corresponding to coupling $0$, to PBC, corresponding to coupling $\beta$, like on the rest of the lattice. This is achieved by defining the system 
on a hypercubic lattice of length $L$ lattice spacings, with the action
\beq
S_{n}[U] = - \frac{\beta}{N_{\scriptscriptstyle{c}}} \sum_{x, \mu \ne \nu}  K_{\mu\nu}^{(n)}(x)\Re\Tr\left[P_{\mu\nu}^{(n)}(x)\right]
\eeq
where $P_{\mu\nu}^{(n)}(x)=U^{(n)}_\mu(x)U^{(n)}_\nu(x+a\hat{\mu}) {U^{(n)}_\mu}^\dagger(x+a\hat{\nu}) {U^{(n)}_\nu}^\dagger(x)$ is the elementary plaquette operator at site $x$ 
on the plane $(\mu,\nu)$, and $K_{\mu\nu}^{(n)}(x)$ is 
a numerical factor used to modify the boundary conditions. It is defined as:
\beq
K_{\mu\nu}^{(n)}(x) = K^{(n)}_\mu(x) K^{(n)}_\nu(x+a\hat{\mu}) K^{(n)}_\mu(x+a\hat{\nu}) K^{(n)}_\nu(x),
\eeq
\beq
K_{\mu}^{(n)}(x) =
\begin{cases}
c(n), \qquad &\mu=1, \qquad x_1=L-a, \qquad 0 \le x_0,\,x_2,\,x_3 < L_{\dd},\\
\\[-1em]
1,    \qquad &\text{elsewhere},
\end{cases}
\eeq
where the size of the
three-dimensional defect is given by $L_\dd \times L_\dd \times L_\dd$, while
$c(n)$ is an \emph{a priori} arbitrary function. The r\^ole of the latter is to interpolate between OBC when $c=0$ to PBC when $c=1$.

Second, an ensemble is obtained from equilibrium MCMC simulation with OBC and the
standard combination of local HB+OR updates. We call $\nbet$ the number of 
full lattice sweeps between successive configurations along the Markov Chain with OBC.
Each configuration in this ensemble is used as starting point for a
non-equilibrium evolution towards a configuration with PBC. In particular,
iterations of the same local updating algorithm are run while
gradually changing the value of $c$ along a so-called \emph{protocol},
\beq
c(n)=1-\frac{n}{\nstep-1}
\eeq
where $\nstep$ is the number of steps separating the two different 
boundary conditions. We remark that in this framework the system does not have to reach equilibrium
between successive steps.

Third, any desired observable is calculated on the configurations of the
newly obtained ensemble with PBC by performing an appropriate 
\emph{reweighting} with the statistical weight $e^{-W}$,
\beq
\label{eq:estimator}
\braket{\mathcal{O}}_{\NE} = \frac{\braket{\mathcal{O}e^{-W}}_{\f}}{\braket{e^{-W}}_{\f}},
\eeq
where $W$ is the \emph{work} spent along a non-equilibrium evolution,
that can be calculated explicitly as follows,
\beq
\label{eq:work}
W = \sum_{n=0}^{\nstep-1} \left\{ S_{n+1}[U_n] - S_{n}[U_n]\right\}~,
\eeq
with $U_n$ the gauge configuration at the $n^{\text{th}}$ step of the out-of-equilibrium transformation. 

This strategy behind Eq.~\eqref{eq:estimator} can be summarized as follows: we aim at sampling
a target distribution $p$ with PBC by reweighting a sample obtained
from the prior distribution $q_0$ with OBC and driven towards PBC through a sequence
of non-equilibrium \emph{forward transitions} with probability
\beq
\mathcal{P}_{\mathrm{f}}[U_0,\dots,U] = \prod_{n=1}^{\nstep} \mathcal{P}_{n} (U_{n-1} \to U_n),
\eeq
where each transition probability $\mathcal{P}_{n}$ is defined by the intermediate action $S_n$.
The expectation value of any observable $\mathcal{O}$ with respect to the target distribution
\beq
p[U] = \frac{1}{Z}e^{-S[U]}, \qquad \qquad S \equiv S_{\nstep},
\eeq
can be computed using Eq.~\eqref{eq:estimator}, where the average over the evolutions is formally defined as
\beq
\langle \dots \rangle_{\mathrm{f}} = \int [\dd U_0 \dots \dd U ] q_0[U_0] \, \mathcal{P}_{\mathrm{f}}[U_0,\dots, U] \, \dots, \qquad 
q_0[U_0] = \frac{1}{Z_0}e^{-S_0[U_0]}.
\eeq
We refer to Ref.~\cite{Bonanno:2024udh} for a more in-depth discussion of these non-equilibrium evolutions.
%

\begin{figure}[!t]
\centering
\includegraphics[scale=1.1]{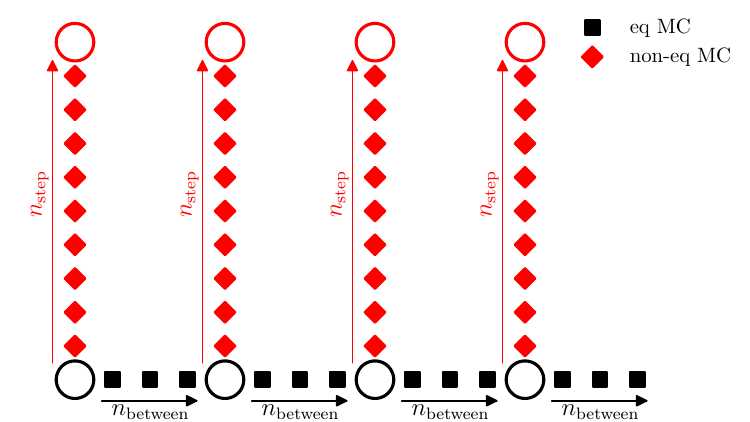}
\caption{Sketch of our out-of-equilibrium Monte Carlo setup.}
\label{fig:sketch}
\end{figure}

A sketch of our out-of-equilibrium Monte Carlo (MC) setup is shown in
Fig.~\ref{fig:sketch}. The horizontal axis represents equilibrium evolutions
with OBCs, used to sample the prior distribution $q_0$ (OBC). The vertical axis
represents out-of-equilibrium evolutions with protocol $c(n)$, used to gradually
reach the target distribution $p$ (PBC). The out-of-equilibrium trajectories are
$\nstep$ update steps long, and are separated by $\nbet$ steps, the
latter performed at equilibrium.

The ratio between the partition functions of the prior $q_0$ and target $p$, $Z_0$ and $Z$ respectively, defines the free energy difference $\Delta F$
\beq
e^{-\Delta F} = \frac{Z}{Z_{0}} \,,
\eeq
which can also be computed from the work along the non-equilibrium evolutions using the celebrated Jarzynski's equality~\cite{Jarzynski:1997ef},
\beq\label{eq:JARZ_eq}
\braket{\exp (-W)}_{\rm f} =  \exp{(-\Delta F)}\,.
\eeq

Of crucial importance is clearly the quality of the sampling
of $p$. This can be evaluated using two figures of merit that estimate how far
from equilibrium the evolution is. The first is the reverse Kullback--Leibler
divergence,
\beq\label{eq:DKL}
\DKL(q_0 \mathcal{P}_{\f} \| p \mathcal{P}_{\r}) = \braket{W}_{\f} - \Delta F \geq 0,
\eeq
which measures the reversibility of the evolution defined by a given protocol $c(n)$, 
and the second is the Effective Sample Size (ESS), estimated with
\beq
\label{eq:ESS_def}
\hat{\ESS} \equiv \frac{\langle e^{-W} \rangle_{\f}^2}{ \langle e^{-2W} \rangle_{\f}} = \frac{1}{\langle e^{-2(W-\Delta F)} \rangle_{\f}}~,
\eeq
The ESS has an interesting intuitive interpretation. It is easy to show
that the variance of $\braket{\mathcal{O}}$ obtained from $n$ non-equilibrium
measurements is related to the corresponding variance sampled from the
target distribution $p$ as follows,
\beq
\label{eq:ESS_var}
\frac{\Var(\mathcal{O})_{\NE}}{n} = \frac{\Var(\mathcal{O})_p }{n \, \ESS}~.
\eeq
Hence, the former is
larger by a factor of $1/\ESS$ with respect to the latter, obtained by sampling
$p$ at equilibrium. 

In the above formula, the autocorrelations between evolutions have been ignored for simplicity.
However, they must be accounted for whenever an estimate of the
total cost of using the above strategy is needed. A possible way of estimating
the latter was provided in Ref.~\cite{Bonanno:2024udh}. We report it here,
\beq\label{eq:variance}
\Var(\mathcal{O})_{\NE} \times (\nstep + \nbet)  \simeq \Var(\mathcal{O})_p \frac{2 \tauint}{\hat\ESS} \times (\nstep + \nbet).
\eeq
Since this quantity can be computed for any other algorithm, it can be
used as a quantitative statistics-independent metric for the efficiency of the
algorithm in computing a given observable with a target precision.

\section{Numerical results}

In this section we report on the preliminary results obtained for the two
figures of merit $\hat{\ESS}$ and $\DKL$ and for the topological susceptibility.
All our results refer to simulations conducted for $\beta=6.40$ on a symmetric
$(L/a)^4 = 30^4$ lattice. Assuming $r_0 \simeq 0.5$ fm for the Sommer scale, this
point corresponds to a lattice spacing $a/r_0\simeq 0.1027(5)$, i.e., $a\simeq
0.05$ fm and $L \simeq 1.5$ fm, which is a large enough box to be insensitive
to finite-size effects in the topological susceptibility with periodic boundary
conditions. The integrated auto-correlation time with PBC of $Q^2$ was estimated to be
$\tauint(Q^2)\simeq 520(20)$ in units of standard Monte Carlo updating steps
using standard algorithms~\cite{Bonati:2015sqt}, see also the caption of
Fig.~\ref{fig:freezing}.

\begin{figure}[!t]
\centering
\begin{subfigure}{.5\textwidth}
\includegraphics[scale=0.49,keepaspectratio=true]{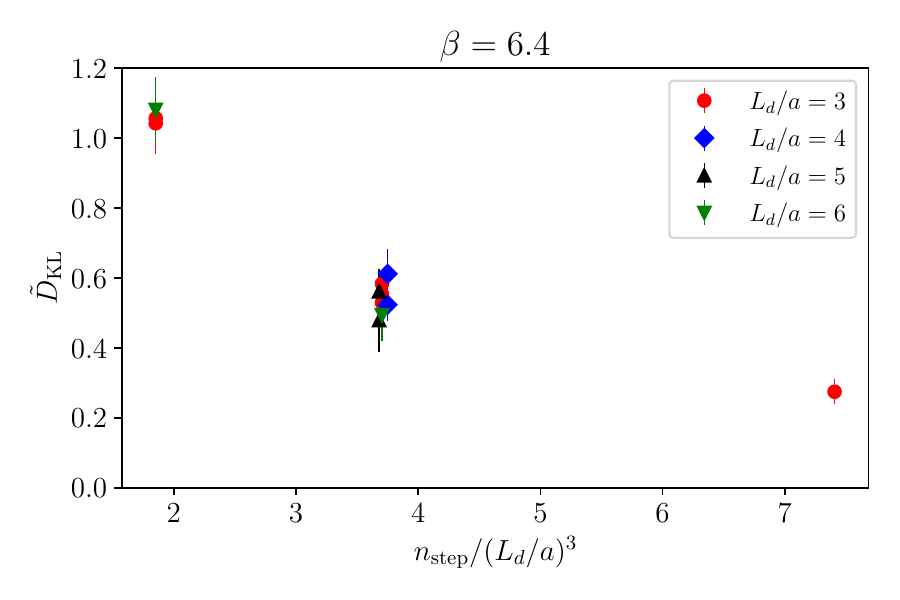}
\end{subfigure}%
\begin{subfigure}{.5\textwidth}
\includegraphics[scale=0.49,keepaspectratio=true]{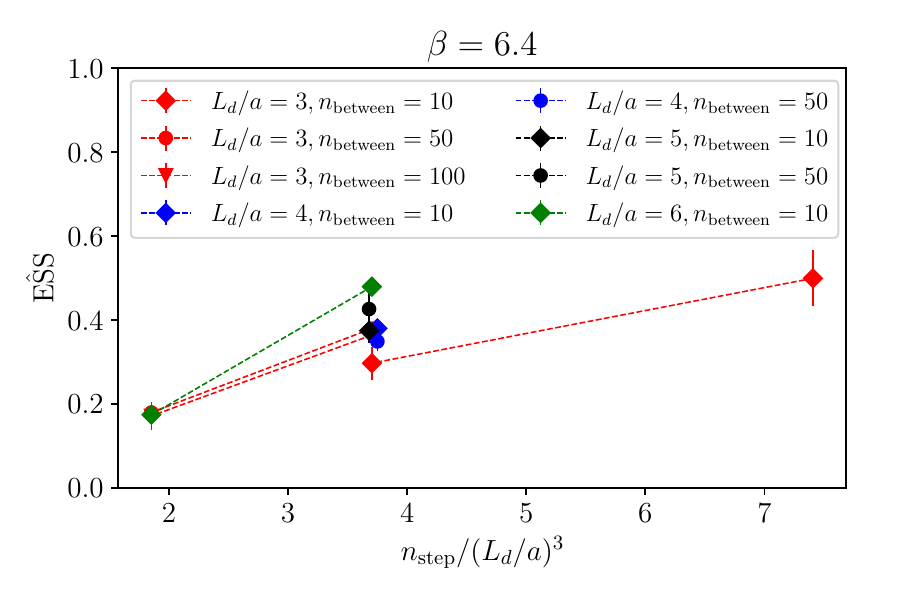}
\end{subfigure}%
\caption{Figures of merit to quantify how far from equilibrium the evolution is as a function of the length of the out-of-equilibrium trajectory $\nstep$ and of the defect size in lattice units $(L_\dd/a)^3$.}
\label{fig:DKL_ESS}
\end{figure}

The values of the figures of $\hat\ESS$ and $\DKL$ can be calculated as follows.
The former is obtained using Eq.~\eqref{eq:ESS_def} on the values of the work defined 
by Eq.~\eqref{eq:work}. The values of $\DKL$ are also obtained from the work by direct computation using Eqs.~\eqref{eq:JARZ_eq} and~\eqref{eq:DKL}.
These quantities were computed for several values of $L_\dd$ and $\nstep$ and 
are displayed in Fig.~\ref{fig:DKL_ESS} as a function of
$\nstep/(L_\dd/a)^3$. In Ref.~\cite{Bonanno:2024udh}, the latter was identified
as the proper scaling variable for these figures of merit in the case of the $2d$
$\CP^{N-1}$. Fig.~\ref{fig:DKL_ESS} shows that this is also the case for the
$4d$ $\SU(3)$ pure lattice gauge theory. Indeed, to a very good degree of approximation,
the values of $\DKL$ and $\hat\ESS$ are seen to collapse on a single curve 
when represented as a function of $\nstep/(L_\dd/a)^3$. 
Verifying this scaling is key to understanding how to tune the simulation
parameters as the lattice spacing is reduced, and to estimate the
cost of a simulation as a function of $\beta$.

\begin{figure}[!t]
\centering
\begin{subfigure}{.5\textwidth}
\includegraphics[scale=0.49,keepaspectratio=true]{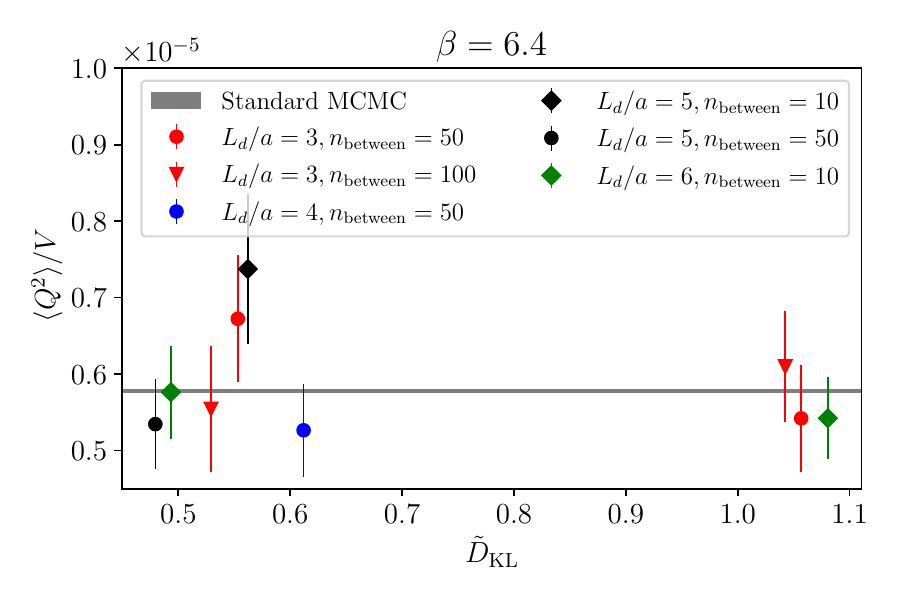}
\end{subfigure}%
\begin{subfigure}{.5\textwidth}
\includegraphics[scale=0.49,keepaspectratio=true]{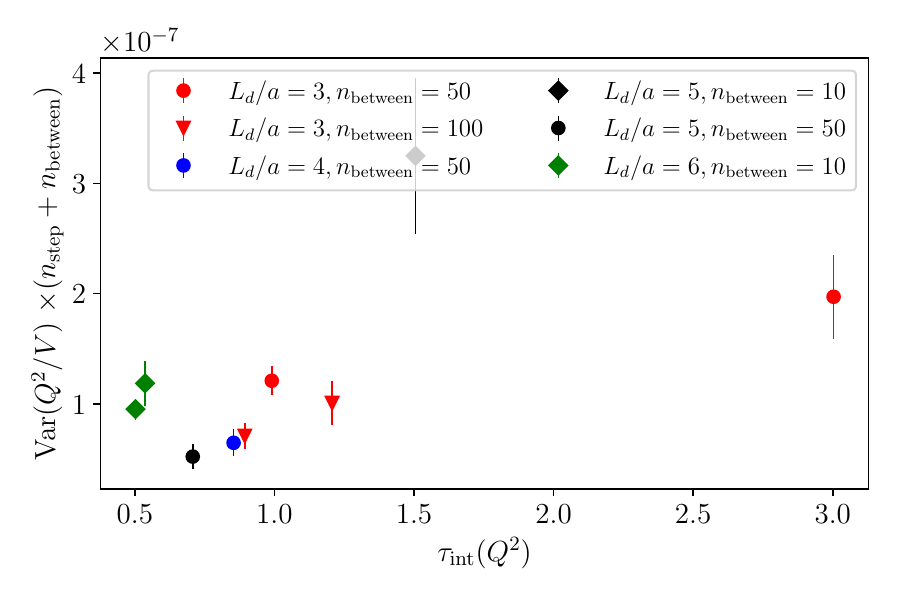}
\end{subfigure}%
\caption{Left: results for the topological susceptibility. Right: results for the variable defined in Eq.~\eqref{eq:variance}, which expresses the efficiency of the out-of-equilibrium algorithm.}
\label{fig:chi}
\end{figure}

We now move on to our evaluation of the topological susceptibility. The
latter was defined as follows,
\beq
a^4\chi \equiv \frac{1}{V}\braket{Q^2},
\eeq
with $V=(L/a)^4$ and where $Q$ was obtained using a standard clover discretization and computed
after 30 cooling steps. We performed several runs with different values of 
$\nstep$ and $L_\dd$. In the left-hand side panel of Fig.~\ref{fig:chi},
we display the non-equilibrium values thus obtained as a function of $\DKL$.
The value of the topological susceptibility obtained with traditional
algorithms, see Ref.~\cite{Bonanno:2023ple}, is represented as a black horizontal
line. In all cases, the non-equilibrium and traditional determinations 
are in agreement within 1.5 standard deviations at most.
This shows that, for the values of $\DKL$ attained in our settings, the 
susceptibility obtained with non-equilibrium methods is not significantly
affected by the details of the simulation and it reproduces the equilibrium 
estimates.

Finally, in order to assess the efficiency of the non-equilibrium method
we have estimated the cost of computing the topological susceptibility with this approach,
using the left-hand side of Eq.~\eqref{eq:variance}. The latter quantity was
computed for each of the available values of $\nstep$ and $L_\dd$ and is
displayed in the right-hand side panel of Fig.~\ref{fig:chi} as a function of
the integrated autocorrelation time $\tauint(Q^2)$. Smaller autocorrelation
times are obtained for larger values of either $L_\dd$ or $\nbet$, or both, as 
expected. In the regime of small autocorrelations, $\tauint(Q^2)\sim 0.5$--1.5, the values of the cost gather
around $1$, which appears to be the optimal regime for our setup. 
Given that at the currently explored value of $\beta$ this was achieved for values of $\nbet$ and $\nstep$ of the order of a few hundreds, this means that the cost of 1 decorrelated out-of-equilibrium trajectory in terms of number of updates only improves by a small factor on the results obtained with the standard algorithm, $\tauint(Q^2)=520(20)$. However, the gain obtained with our method is expected to increase at larger values of $\beta$, owing to the improved scaling of $\tauint(Q^2)$ obtained with OBC.

\section{Conclusions and future outlooks}

In this proceeding we presented a preliminary study of the topological
susceptibility of the pure $\SU(3)$ lattice gauge theory using out-of-equilibrium
simulations. We implemented the same setup already proposed in and tested in
Ref.~\cite{Bonanno:2024udh}, and our results are encouraging. The figures of merit
$\hat\ESS$ and $\DKL$ are found to only depend on the scaling variable
$\nstep/(L_\dd/a)^3$ rather than on $\nstep$ and $L_\dd$ separately. Moreover, 
our estimate of the topological susceptibility is statistically compatible with
the estimates present in the literature, with an integrated auto-correlation time 
for $Q^2$ that is slightly smaller than the one attained with traditional algorithms.

In the near future we plan to extend the present preliminary study in two different
directions. Firstly, by probing larger values of $\beta$ and by studying
systematically how the cost of
the out-of-equilibrium behaves as the lattice spacing is reduced.
Secondly and most importantly, by combining the setup 
discussed and described above with the discrete neural network layers
composing Normalizing Flows. Stochastic Normalizing Flows fit naturally
in the out-of-equilibrium setup, and have the potential of improving
the efficiency of our approach even further, see for example the improvement obtained for $\SU(3)$ gauge theory in Ref.~\cite{Bulgarelli:2024cqc}. This would be achieved by reducing
the amount of updating steps necessary to obtain a fixed $\ESS$, at the price of performing a once-and-for-all training procedure with moderate costs.

\vspace*{-0.5\baselineskip}
\acknowledgments
We thank L.~Giusti, A.~Patella and F.~Sanfilippo for insightful comments and discussions. Numerical simulations have been performed on the DiRAC Data Intensive service at Cambridge. The work of C.~Bonanno is supported by the Spanish Research Agency (Agencia Estatal de Investigación) through the grant IFT Centro de Excelencia Severo Ochoa CEX2020-001007-S and, partially, by grant PID2021-127526NB-I00, both funded by MCIN/AEI/10.13039/ 501100011033. A.~Nada acknowledges support by the Simons Foundation grant 994300 (Simons Collaboration on Confinement and QCD Strings), from the European Union - Next Generation EU, Mission 4 Component 1, CUP D53D23002970006, under the Italian PRIN “Progetti di Ricerca di Rilevante Interesse Nazionale – Bando 2022” prot. 2022TJFCYB and from the SFT Scientific Initiative of INFN. 

\bibliographystyle{JHEP}
\bibliography{biblio}

\end{document}